\documentclass[onecolumn,preprintnumbers,amsmath,amssymb,showkeys,12ft]{revtex4}
\usepackage{graphicx}% Include figure files
\usepackage{dcolumn}% Align table columns on decimal point
\usepackage{bm}% bold math
\begin{document}
\draft
\title{Gibbs' theorem for open systems with incomplete statistics}
\author{G. B. Ba\u{g}c\i}
\email{baris.bagci@ege.edu.tr}
\address {Department of Physics, Faculty of Science, Ege University, 35100 Izmir, Turkey}

\pagenumbering{arabic}

\begin{abstract}
Gibbs' theorem, which is originally intended for canonical ensembles
with complete statistics has been generalized to open systems with
incomplete statistics. As a result of this generalization, it is
shown that the stationary equilibrium distribution of inverse power
law form associated with the incomplete statistics has maximum
entropy even for open systems with energy or matter influx. The
renormalized entropy definition given in this paper can also serve
as a measure of self-organization in open systems described by
incomplete statistics.
\end{abstract}

\newpage \setcounter{page}{1}

\maketitle

\section{\protect\bigskip Introduction}

Many different entropy measures such as R\'{e}nyi [1], Sharma-Mittal
[2] and Tsallis measures [3] have been proposed in order to
generalize Boltzmann-Gibbs (BG) entropy. Despite their differences
though, all these entropy measures share a common feature in that
they are all based on the assumption of complete statistics, which
is even shared by the entropy measure they claim to generalize i.e.,
BG entropy. The assumption of complete statistics implies that all
states regarding the system is countable and known completely by us
so that we have full knowledge of the interactions taking place in
the system of interest, thereby implying the ordinary normalization
condition $\sum_{i}p_{i}=1$. However, this scenario is challenged in
some cases of interest since one cannot obtain all the information
regarding the system under investigation in these particular
situations. One such possibility is fractal phase spaces [4, 5], since one can then have
singularities and inaccessible points, which invalidates the assumption
of complete statistics.

Recently, Wang [6] proposed a new nonadditive entropy based on
Tsallis entropy by replacing the complete normalization condition
with an incomplete one depending on a free positive parameter $q$.
This new framework is called the formalism of incomplete statistics
(IS). The parameter $q$ plays the role of Hausdorff dimension
divided by the topological dimension of the phase space when IS
formalism is applied to nonequilibrium systems evolving in
hierarchically heterogeneous phase space, connecting the concept of
information and the topological dimension of some fractal sets as
pointed out earlier by El Naschie [7]. However, the general physical
interpretation of this parameter is related to the neglected
interaction. When no interaction is neglected, the parameter $q$
becomes equal to unity, thereby reducing the results of IS formalism
to those of complete statistics described by BG entropy.

Despite the simplicity of the proposal by Wang and the treatments of
some important issues in this framework such as zeroth law of
thermodynamics [8] and the IS formalism with different $q$ indices
[9], there are still open problems in IS formalism like the
definition of physical temperature [10, 11]. The aim of this paper is to clarify another such issue concerning IS formalism,
namely, its applicability to open systems and the ordering of the
entropies.

The Gibbs' theorem states that the canonical equilibrium
distribution, of all the normalized distributions having the same
mean energy, is the one with maximum entropy. However, the Gibbs'
theorem rests on two assumptions. Firstly, the stationary
equilibrium distribution, being canonical, is of exponential form.
Secondly, Gibbs assumed that all the compared distributions have the
same mean energy values. The former assumption has recently been challenged
by Abe and Rajagopal [12]. Although they originally aimed
their generalization for Tsallis entropy, their treatment can be
considered valid also for IS formalism, since Abe and Rajagopal
relied on the use of stationary equilibrium distributions of the
q-exponential form, which are common for both Tsallis entropy and IS
formalism. However, these authors only considered the
generalization of the Gibbs' theorem, assuming the equality of the
mean energies of the distinct distributions.

The generalization of Gibbs' theorem to open systems was first made by Klimontovich and is called the S-theorem [13-17]. Open systems are those in which the average internal energy is not constant because of the interactions with the environment. It is in this sense that the Gibbs' theorem needs to be generalized, since Gibbs assumed the equality of the
mean energies of the distributions. Unfortunately, Klimontovich's
treatment is based on the use of BG entropy, which in turn is based
on the assumption of complete statistics and yields to stationary
equilibrium distributions of exponential form.

In this paper, we generalize Gibbs' theorem for
open systems described by incomplete formalism. We show that the
stationary equilibrium distribution of inverse power law form
associated with IS formalism has maximum entropy even in open
systems with energy or matter influx. The general derivation here can be used as a criterion for self-organization in open systems with incomplete statistics.

The paper is organized as follows: In Section II, we give  a
heuristic review of IS formalism. The generalization of Gibbs'
theorem for open systems with incomplete statistics is given in
Section III. The conclusions is given in Seection IV.

\section{Incomplete statistics}

The point of departure of IS formalism is the nonadditive Tsallis
entropy [3]

\begin{equation}
S_{q}^{Tsallis}(p)=\frac{\sum_{i}^{W}p_{i}^{q}-\sum_{i}^{W}p_{i}}{1-q},
\end{equation}

\noindent where $p_{i }$ is the probability of the system in
the $i$th microstate and $W$ is the total number of the configurations of
the system. The Boltzmann constant $k$ is set to unity
throughout the paper. The entropic index $q$ is a real number, which
characterizes the degree of nonadditivity since Tsallis entropy
obeys the following pseudo-additivity rule:

\begin{equation}
S_{q}(A+B)/k=[S_{q}(A)]+[S_{q}(B)]+(1-q)[S_{q}(A)][S_{q}(B)],
\end{equation}

\noindent where A and B are two independent systems with p$_{ij}$(A+B)$=$p$_{i}(A)$p$%
_{j}(B)$. Obviously, this formalism assumes that we have complete
access to all random variables since the summation in Eq. (1) is
over the total number of configurations $W$. This assumption also
lies at the center of the normalization of probability distribution,
since it considers that all the information relevant to the physical
system under consideration is accessible. On the other hand,
IS formalism challenges this issue since a complete description of a
physical system requires the knowledge of the exact Hamiltonian and
the exact solution of its corresponding equations of motion. However, it is possible in practice that we may not know analytically all the interactions, which must appear in the
relevant Hamiltonian. This failure in our knowledge of the exact
Hamiltonian results in the incompleteness of the countable states.
If this is the case, the ordinary normalization of the probability
distribution will not hold and must be avoided. This observation
lies at the heart of IS formalism, since Wang proposes to replace the
usual i.e., complete normalization $\sum_{i}^{W}p_{i}=1$ with the
incomplete normalization given by

\begin{equation}
\sum_{i}^{w}p_{i}^{q}=1,
\end{equation}

\noindent where $w$ denotes the states accessible to us. Since these states do
not now form a complete set, $w$ can be greater or smaller than the
real number of all possible states. The incompleteness parameter $q$
is positive and can be considered as a measure of neglected
interactions. When all the interactions are taken into account, it is
equal to 1, recovering the ordinary normalization. As a result of
incomplete normalization, the expectation value of an observable
$\hat{O}$ is consistently given by

\begin{equation}
\left\langle \hat{O}\right\rangle =\sum_{i=1}^{w}p_{i}^{q}O_{i}.
\end{equation}

The incomplete normalization given by Eq. (3) enables one to write a
new entropy based on incomplete statistics whose point of departure
is Tsallis entropy in Eq. (1). This new entropy expression reads

\begin{equation}
S_{q}(p)=\frac{1-\sum_{i}^{w}p_{i}}{1-q}.
\end{equation}

\noindent The stationary equilibrium distribution for IS entropy in the
canonical case can be found by applying the Lagrange method

\begin{equation}
\delta (S_{q}+\frac{\alpha
}{1-q}\sum\limits_{i=1}^{w}p_{i}^{q}-\alpha \beta
\sum_{i=1}^{w}p_{i}^{q}\varepsilon _{i})=0.
\end{equation}

The resulting canonical equilibrium distribution [6] in IS formalism
is then given by

\begin{equation}
p^{eq}_{i}=[1-(1-q)\beta \varepsilon _{i}]^{1/(1-q)},
\end{equation}

\noindent apart from normalization. The energy of the $i$th microstate is
denoted by $\varepsilon _{i}$ where the Lagrange multipliers are
denoted by $\alpha$ and $\beta$. However, it is easy to verify that
the canonical equilibrium distribution given by Eq. (7) is not
invariant under the uniform translation of the energy parameter
$\varepsilon _{i}$. Moreover, the Lagrange multiplier $\beta$ is not
identical with the inverse temperature [10, 11].

These difficulties are easily overcome by the maximization of the IS
entropy in Eq. (5) as follows

\begin{equation}
\delta (S_{q}-\alpha\sum\limits_{i=1}^{w}p_{i}^{q}-\beta
\sum_{i=1}^{w}p_{i}^{q}\varepsilon _{i})=0,
\end{equation}

\noindent which results

\begin{equation}
p_{i}^{eq}=[1-(1-q)q\beta (\varepsilon
_{i}-U_{q})/\sum\limits_{j=1}^{w}p_{j}]^{1/(1-q)},
\end{equation}

\noindent where $U_{q}=\sum_{i=1}^{w}p_{i}^{q}\epsilon _{i}$ is the internal
energy and the normalization constant is not explicitly written.
This canonical equilibrium distribution is invariant under uniform
translation of energy parameter and the Lagrange multiplier $\beta$
is identical with inverse temperature [11].

\section{Open systems with Incomplete statistics}

The Gibbs' theorem states that the canonical equilibrium
distribution has the maximum entropy compared to any other
distribution with the same mean energy. Therefore, it can be
considered as an indication of the importance of the canonical
ensemble since it is the state with maximum entropy. However, Gibbs'
theorem relies on two major assumptions. Firstly, the system under
consideration is governed by complete statistics, which is described
by BG measure. Due to this assumption, Gibbs' theorem is limited to
the cases where the stationary equilibrium distribution is
exponential. Secondly, Gibbs' theorem is limited to the cases where
the average internal energy is kept constant. It is our aim in this
paper to generalize Gibbs' theorem for open systems with incomplete
statistics. It is highly probable indeed that one cannot write all
the interactions governing such a system, since open systems are
subject to many interactions. Therefore, their equations of motion are not fully
solvable.

This in turn will result in not all states being
accessible to us, justifying the use of incomplete statistics. In
addition, an open system might have a metastable stationary state
described by an inverse power law as IS suggests instead of an
exponential as Klimontovich assumed [13-15]. These considerations
force us to generalize Gibbs' theorem for open systems with
incomplete statistics. In order to do this, we first define a new
quantity named renormalized entropy $R_{q}^{IS}$ as

\begin{equation}
R_{q}^{IS}\equiv\
S_{q}^{neq}(r)-\widetilde{S}_{q}^{eq}(\widetilde{p}_{eq}).
\end{equation}

\noindent The generalization of Gibbs' theorem is now equivalent to showing
that renormalized entropy expression in Eq. (10) is negative i.e.,
$R_{q}^{IS}<0$, since this implies that
$\widetilde{S}_{q}^{eq}>S_{q}^{neq}$. However, we know that this
cannot be proved on the basis of ordinary Gibbs' theorem, since it
assumes that the internal energy is kept constant, which is not the
case for open systems. Therefore, Klimontovich generalized Gibbs' theorem for open systems with complete statistics by equating the mean energies of the equilibrium and
nonequilibrium states [13-15]. Due to this equalization of the
effective mean energy, we denote the equilibrium entropy by a tilde
since this is not the original equilibrium entropy but the one
obtained after the effective mean energy equalization. Two distinct
incomplete probability distributions i.e. $\widetilde{p}_{eq}$ and
$r$ in Eq. (10) denotes the renormalized equilibrium and
nonequilibrium probability distributions respectively. The
corresponding IS entropy expressions are denoted by
$\widetilde{S}_{q}^{eq}(\widetilde{p}_{eq})$ and $S_{q}^{neq}(r)$.
From now on, we will drop the subscript from the equilibrium
probability distribution so that it should be understood that the
probability distributions $p$ and $\widetilde{p}$ denote the
ordinary and renormalized equilibrium distributions, respectively.
The renormalized equilibrium probability distribution
$\widetilde{p}$ and nonequilibrium probability distribution $r$ obey
the incomplete normalization summarized by Eq. (3) i.e.,
$\sum_{i}^{w}\widetilde{p}_{i}^{q}=\sum_{i}^{w}r_{i}^{q}=1$, which
is implicity taken into account in the definition of IS entropy
given by Eq. (5).

In order to proceed, we need to define effective
mean energy in terms of the equilibrium state associated with the
incomplete statistics. This can be achieved by defining

\begin{equation}
U_{eff}\equiv-\ln _{q}p_{i},
\end{equation}

\noindent where the $q$-logarithm is simply defined as

\begin{equation}
\ln _{q}(x)=\frac{x^{1-q}-1}{1-q}.
\end{equation}

\noindent This definition of effective mean energy is central to our
generalization and therefore requires some explanation. The
effective mean energy is defined in terms of the unnormalized
equilibrium distribution. In this sense, if we apply this definition
to the unnormalized equilibrium distribution given by Eq. (7),
we see that $U_{eff}=\beta\varepsilon_{i}$. The application of the
effective mean energy to the canonical equilibrium distribution in
Eq. (9) however results in $U_{eff}=q\beta(\varepsilon_{i}-U_{q})$.
This observation explains why it is called effective mean energy
since it is always proportional to the multiplication of the
Lagrange multiplier $\beta$ associated with the internal energy
constraint and the energy of the $i$th microstate. The calculations in
this paper are general in the sense that both equilibrium
distributions can be used although a consistent treatment would be
through the adoption of the canonical equilibrium distribution in
Eq. (9) due to its explicit dependence on temperature through
$\beta$. The open systems are usually treated by using a control
parameter, which controls the matter or energy influx into the
system due to its interaction with the environment. In this sense,
the state with the zero value of the control parameter is the
equilibrium distribution and all the other stationary states with
control parameter values different than zero correspond to
nonequilibrium distributions.

Having clarified this important issue,
we can rewrite the equalization of effective mean energies of the
two states as

\begin{equation}
\left\langle U_{eff}\right\rangle ^{(req)}=\left\langle
U_{eff}\right\rangle ^{(neq)},
\end{equation}

\noindent where superscripts $(req)$ and $(neq)$ denote the renormalized
equilibrium and ordinary nonequilibrium states, respectively. The corresponding averages
must be taken in terms of $\widetilde{p}_{i}^{q}$ and $r_{i}^{q}$. The Eq. (13) can be explicitly written as

\begin{equation}
\sum_{i=1}^{w}\widetilde{p}_{i}^{q}U_{eff}=\sum_{i=1}^{w}r_{i}^{q}U_{eff}.
\end{equation}

\noindent The substitution of the effective mean energy defined in Eq. (11)
into Eq. (14) yields

\begin{equation}
\sum_{i=1}^{w}\widetilde{p}_{i}^{q}(\frac{p_{i}^{1-q}-1}{q-1})=\sum_{i=1}^{w}r_{i}^{q}(%
\frac{p_{i}^{1-q}-1}{q-1}).
\end{equation}

\noindent Due to the normalization i.e.,
$\sum_{i}^{w}\widetilde{p}_{i}^{q}=\sum_{i}^{w}r_{i}^{q}=1$, the
above equation can be rewritten as

\begin{equation}
\sum_{i=1}^{w}\widetilde{p}_{i}^{q}p_{i}^{1-q}=\sum_{i=1}^{w}r_{i}^{q}p_{i}^{1-q}.
\end{equation}

\noindent The probability distribution $\widetilde{p}$ can be considered as
the normalized and renormalized (i.e., effective mean energy
equalization) counterpart of the ordinary equilibrium distribution
$p$. Therefore, we can substitute
$\widetilde{p}_{i}^{q}=\frac{p_{i}^{q}}{\sum\limits_{i}p_{i}^{q}}$
into Eq. (16) to obtain

\begin{equation}
\sum_{i=1}^{w}\widetilde{p}_{i}=\sum_{i=1}^{w}r_{i}^{q}\widetilde{p}_{i}^{1-q}.
\end{equation}

\noindent On the other hand, we can also obtain an explicit form of
renormalized entropy defined by Eq. (10) by substituting the IS
entropy in Eq. (5) explicitly, which gives

\begin{equation}
R_{q}^{IS}=\frac{1}{(q-1)}(\sum_{i=1}^{w}r_{i}-\sum_{i=1}^{w}\widetilde{p}_{i}).
\end{equation}

\noindent Making use of the equalization of the effective mean energies of
equilibrium and nonequilibrium states given by Eq. (17), we obtain

\begin{equation}
R_{q}^{IS}=\frac{1}{(q-1)}(\sum_{i=1}^{w}r_{i}-\sum_{i=1}^{w}r_{i}^{q}\widetilde{p}_{i}^{1-q}).
\end{equation}

\noindent The final step in our treatment is to show that the renormalized
entropy $R_{q}^{IS}$ is negative for all positive values of the
incompleteness parameter $q$. This can be achieved by rewriting the
above renormalized entropy expression as

\begin{equation}
R_{q}^{IS}=-\sum_{i=1}^{w}r_{i}[\frac{(r_{i}/\widetilde{p}{i})^{q-1}-1}{q-1}].
\end{equation}

\noindent Since r$_{i}/p_{i}\geq0$, the following mathematical inequality can
be used [18]

\begin{equation}
\frac{(r_{i}/\widetilde{p}_{i})^{q-1}-1}{q-1}\geq
1-\widetilde{p}_{i}/r_{i}, \qquad q>0.
\end{equation}

\noindent Multiplying both sides of the above inequality with $r_{i}$ and
summing over $i$, we obtain

\begin{equation}
\sum_{i=1}^{w}r_{i}\frac{(r_{i}/\widetilde{p}_{i})^{q-1}-1}{q-1}\geq
\sum_{i=1}^{w}(r_{i}-\widetilde{p}_{i}), \qquad q>0.
\end{equation}

\noindent Comparing the inequality above with the expressions given by Eqs.
(18) and (20), we see that the above inequality takes the form

\begin{equation}
-R_{q}^{IS}\geq (q-1)R_{q}^{IS}, \qquad q>0.
\end{equation}

\noindent Since the above inequality is valid only for $q$ values greater than
zero, it implies

\begin{equation}
R_{q}^{IS}\leq 0.
\end{equation}

\noindent The equality holds only if the two distributions are the same. Since
we assume that the states under question are two different states,
one being renormalized equilibrium state and the other being
nonequilibrium state, we can drop the equality sign above. Moreover,
remembering the original definition of renormalized entropy in Eq.
(10), we see that

\begin{equation}
R_{q}^{IS}=S_{q}^{neq}-\widetilde{S}_{q}^{eq}<0\Rightarrow
\widetilde{S}_{q}^{eq}>S_{q}^{neq},
\end{equation}

\noindent thus the (renormalized) equilibrium entropy is greater than the
nonequilibrium entropy for open systems with incomplete statistics.
Naturally, we recover the result based on complete statistics of BG
entropy i.e., $S^{eq}>S^{neq}$ by taking the $q\rightarrow1$ limit
in Eq. (25).

In summary, we have a more ordered state as the
control parameter increases causing the system to recede away from
equilibrium. This decrease of entropy on ordering is called
self-organization by Haken [19]. Therefore, the renormalized entropy
$R_{q}^{IS}$ can also be taken as a measure of self-organization for
open systems described by  IS.

\section{Conclusions}

By generalizing Gibbs' theorem for open systems described by IS
formalism, we have shown that the stationary equilibrium state
obtained from IS is the state of maximum entropy even in the
presence of energy or matter influx. The treatment here can be
considered as a generalization of ordinary S-theorem (and also of Gibbs'
theorem) developed by Klimontovich, since the latter
is a particular case of the former in the $q\rightarrow1$ limit. In
this sense, the incompleteness seems a more general framework since
the results assuming complete statistics can be obtained from IS
even in the case of open systems. This generalization of Gibbs'
theorem to IS is necessary, since it is less likely that we will have complete knowledge
of the system as the interaction terms governing the system increase
as it would generally be the case with open systems. Finally, we
have shown that one obtains a more ordered state i.e., a state of
lesser entropy as the control parameter increases. In this sense,
the renormalized entropy expression obtained in this paper can serve
as a criterion of self-organization [19] in open systems described
by IS.

\section{ACKNOWLEDGEMENTS}

I thank Professors Q. A. Wang and Donald H. Kobe for valuable
suggestions. I am also grateful to University of North Texas in Denton/Texas for their hospitality during my stay.

\end{document}